\documentclass[12pt,a4paper]{article}
\usepackage[latin1]{inputenc}
\usepackage{amsmath}
\usepackage{amsfonts}
\usepackage{amssymb}
\usepackage[left=2cm,right=2cm,top=2cm,bottom=2cm]{geometry}
\usepackage{tabularx}
\usepackage{subfig}
\usepackage{wrapfig}
\usepackage{tabularx}
\usepackage{sidecap}
\usepackage{graphicx}
\RequirePackage[colorlinks,citecolor=blue,urlcolor=blue,linkcolor=blue]{hyperref}
\begin{document}
\title{Singularity free analysis of a self-similar model of proton structure function at small \textit{x}}
\author{Baishali Saikia$^{1,a}$ and D. K. Choudhury$^{1,2}$ \\  $^1$Department of Physics,\ Gauhati University,
Guwahati-781 014, India.\\$^2$ Physics Academy of North-East, Guwahati-781014, India. \\ $^a$email: baishalipiks@gmail.com}
\date{}
\maketitle

\begin{abstract}
In this paper we make re-analysis of a self-similarity based model of the proton structure function at small \textit{x} pursued in  recent years. The additional assumption is that it should be singularity free in the entire kinematic range $0<\textit{x}<1$. Our analysis indicates that the model is valid in a more restrictive range of $Q^{2}$. We also discuss the possibility of incorporation of Froissart saturation condition in the model.
\end{abstract}

\section{Introduction}
\label{intro}
\label{A}

Self-similarity is a possible feature of multi-partons inside a proton at small \textit{x} suggested by Lastovicka \cite{Last} and pursued by us in recent years \cite{DK1,DK2,DK3,DK4,DK5}. More recently, we have examined its consequences in \cite{DK6} double parton distribution function(dPDF) to be tested at LHC, in Froissart saturation \cite{DK7} and Longitudinal structure function \cite{DK8}.

One of the theoretical limitations of the model of Ref \cite{Last} is that it has a singularity at $x \sim 1.172\times10^{-4}$ which is within the physically allowed range $0<x<1$. Even though outside its phenomenological range of validity $6.2\times10^{-7}<x<0.01$, such singularity is beyond common expectations from any physically viable model of proton structure function $F_2(x,Q^2)$. 

In the present paper, we therefore make a re-analysis of the model of Ref\cite{Last}, demanding it to be singularity free in the entire $x$-range of $0<x<1$.

Present work will also study the consequences of the suggestion of Ref\cite{DK7} in the context of the reported reanalysis. In section\ref{B}, we outline the formalism and in section\ref{C}, we discuss the results. Section\ref{D} contains the conclusions.

\section{Formalism}
\label{B}
\subsection{Proton Structure Function based on self-similarity}
The self-similarity based model of the proton structure function of Ref\cite{Last} is based on transverse momentum dependent parton distribution function(TMD) $f_i(x,k_t^2)$. Here $\textit{k}_t^2$ is the parton transverse momentum squared. Choosing the magnification factors $\left(\frac{1}{x}\right)$ and $\left(1+\frac{k_t^2}{Q_0^2}\right)$, it is written as \cite{Last,DK6,DK7}
\begin{equation}
\label{E1}
log[M^2.f_i(x,k_t^2)]= D_1.log\frac{1}{x}.log\left(1+\frac{k_t^2}{Q_0^2}\right)+D_2.log\frac{1}{x}
+D_3.log\left(1+\frac{k_t^2}{Q_0^2}\right)+D_0^i
\end{equation}

where i denotes a quark flavor. Here $D_1,\ D_2,\ D_3$ are the three flavor independent model parameters while $D_0^i$ is the only flavor dependent normalization constant. M$^2$(=1 GeV$^2$) is introduced to make (PDF) $q_i(x,Q^2)$ as defined below (in Eqn(\ref{E2})) dimensionless. The integral quark densities then defined as:
\begin{equation}
\label{E2}
q_i(x,Q^2) = \int_0^{Q^2}f_i(x,k_t^2)dk_t^2
\end{equation}

As a result, the following analytical parametrization of a quark density is obtained by using Eqn(\ref{E2}) \cite{DK5}
\begin{equation}
\label{E3}
q_i(x,Q^2) = e^{D_0^i}f(x,Q^2)
\end{equation}
where,
\begin{equation}
\label{E4}
f(x,Q^2)= \frac{Q_0^2 \ x^{-D_2}}{M^2\left(1+D_3+D_1log\left(\frac{1}{x}\right)\right)} 
\\
\times \left(\left(\frac{1}{x}\right)^{D_1log \left(1+\frac{Q^2}{Q_0^2}\right)} \left(1+\frac{Q^2}{Q_0^2}\right)^{D_3+1}-1 \right)
\end{equation}
is flavor independent. Using Eqn(\ref{E3}) in the usual definition of the structure function $F_2(x,Q^2)$, one can get
\begin{equation}
\label{E5}
F_2(x,Q^2)=x\sum_i e_i^2 \left( q_i(x,Q^2)+ \bar{q}_i(x,Q^2)\right) 
\end{equation}
or it can be written as
\begin{equation}
\label{E6}
F_2(x,Q^2)=e^{D_0}xf(x,Q^2)
\end{equation}

where $D_0=\sum_i D_0^i$. From HERA data \cite{H1,ZE}, Eqn(\ref{E6}) was fitted with
\begin{eqnarray}
\label{E7}
D_0 &=& 0.339\pm 0.145 \nonumber \\
D_1 &=& 0.073\pm 0.001 \nonumber \\
D_2 &=& 1.013\pm 0.01 \nonumber \\
D_3 &=& -1.287\pm 0.01 \nonumber \\
Q_0^2 &=& 0.062\pm 0.01 \ GeV^2
\end{eqnarray}
in the kinematical region,
\begin{eqnarray}
\label{E8}
0.62\times10^{-7}\leq x\leq 10^{-2} \nonumber \\
0.045\leq Q^2 \leq 120 \ GeV^2
\end{eqnarray}

\subsection{Singularity free Structure Function}
The model of Ref\cite{Last} has two inherent limitations. First, the parameter $D_3$ is negative contrary to the expectations of positivity of the fractal dimensions \cite{DK1}. Secondly, due to its negative value, Eqn(\ref{E6}) develops a singularity at $\textit{x}\backsim1.172\times10^{-4}$ \cite{DK5} as it satisfies the condition $1+D_3+D_1log\frac{1}{x}=0$, contrary to the expectation of a physically viable form of Structure Function.

The model can be made singularity free under following specific conditions of the model parameters.\\
\\
\textbf{Case 1 :} If $D_1, D_3\ll D_2$ in Eqn(\ref{E1}) then the PDF Eqn(\ref{E3}) and the Structure Function Eqn(\ref{E6}) will be of the form:
\begin{equation}
\label{E9}
q_i(x,Q^2)=\frac{e^{D_0^i}\ Q^2\ x^{-D_2}}{M^2}
\end{equation}
\begin{equation}
\label{E10}
F_2(x,Q^2)=\frac{e^{D_0}\ Q^2\ x^{-D_2+1}}{M^2}
\end{equation}\\
\\
\textbf{Case 2 :} In this case $D_1\ll D_2, D_3$ in Eqn(\ref{E1}) then the corresponding expressions for the PDF and Structure Function in this limit are respectively: 
\begin{equation}
\label{E11}
q_i(x,Q^2)=\frac{e^{D_0^i}\ Q_0^2\ x^{-D_2}}{M^2 \left(1+D_3\right)}\ \left(\left(1+\frac{Q^2}{Q_0^2}\right)^{D_3+1}-1\right)
\end{equation}
\begin{equation}
\label{E12}
F_2(x,Q^2)=\frac{e^{D_0}\ Q_0^2\ x^{-D_2+1}}{M^2 \left(1+D_3\right)}\ \left(\left(1+\frac{Q^2}{Q_0^2}\right)^{D_3+1}-1\right) 
\end{equation}\\
\\
\textbf{Case 3 :} In this case, $D_3\ll D_1, D_2$ in Eqn(\ref{E1}) then the corresponding PDF and the Structure Function are set in the form:
\begin{equation}
\label{E13}
q_i(x,Q^2)=\frac{e^{D_0^i}\ Q^2\ x^{-D_2}}{M^2 \left(1+D_1log\frac{1}{x}\right)}
\\
\times \left( \left(\frac{1}{x}\right)^{D_1log\left(1+\frac{Q^2}{Q_0^2}\right)}\left(1+\frac{Q^2}{Q_0^2}\right)-1\right)
\end{equation}
\begin{equation}
\label{E14}
F_2(x,Q^2)=\frac{e^{D_0}\ Q^2\ x^{-D_2+1}}{M^2 \left(1+D_1log\frac{1}{x}\right)}
\\
\times \left(\left(\frac{1}
{x}\right)^{D_1log\left(1+\frac{Q^2}{Q_0^2}\right)}\left(1+\frac{Q^2}{Q_0^2}\right)-1\right)
\end{equation}
respectively.\\
\\
\textbf{Case 4 :} This is the most general case for the singularity free model, Eqn(\ref{E6}) under the condition that $D_1, D_2, D_3$ are positive.

\subsection{Froissart inspired Proton structure function based on self-similarity}
It is to be noted that the variable in which the supposed fractal scaling in the quark distribution occur is not known for the underlying theory. In Ref\cite{Last}, the choice of $\frac{1}{x}$ as one of the magnification factors was taken presumably because of the power law form of the quark distribution at small $x$ found in Gl{\"u}ck-Reya-Vogt (GRV) distribution \cite{GRV}. However, this form is not derived for the theory but rather inspired by the power law distribution in $\textit{x}$ assumed for GRV distribution for QCD evaluation. The choice of $\frac{1}{x}$ as the proper scaling variable is not an established result of the underlying theory.

A more plausible variable appears instead to be $log\frac{1}{x}$ \cite{BL} which confirms to the Froissart saturation \cite{FR,ROY} of high energy interaction.

With this choice, Transverse Momentum Dependent Structure Function (TMD) and Parton Distribution Function (PDF) now take the following forms \cite{DK7}
\begin{equation}
\label{E15}
log\left(M^2.\tilde{f}_i(x,k_t^2)\right)= \tilde{D}_1log\left(log\left(\frac{1}{x}\right)\right).log\left(1+\frac{k_t^2}{\tilde{Q}_0^2}\right)
\\
+\tilde{D}_2log\left(log\left(\frac{1}{x}\right)\right)+\tilde{D}_3log\left(1+\frac{k_t^2}{\tilde{Q}_0^2}\right)+\tilde{D}_0^i
\end{equation}
and
\begin{equation}
\label{E16}
\tilde{q}_i(x,Q^2)= e^{\tilde{D}_0^i}\tilde{f}(x,Q^2)
\end{equation}

where,
\begin{equation}
\label{E17}
\tilde{f}(x,Q^2)=\frac{\tilde{Q}_0^2 \ x^{-\tilde{D}_2}}{M^2\left(1+\tilde{D}_3+\tilde{D}_1log\left(log\left(\frac{1}{x}\right)\right)\right)}
\\
\times \left\lbrace log\left(log\left(\frac{1}{x}\right)\right)^{\tilde{D}_1 log\left(1+\frac{Q^2}{\tilde{Q}_0^2}\right)}\left(1+\frac{Q^2}{\tilde{Q}_0^2}\right)^{\tilde{D}_3+1}-1\right\rbrace 
\end{equation}
The structure function as defined in Eqn(\ref{E6}) becomes
\begin{equation}
\label{E18}
F_2(x,Q^2)= e^{\tilde{D_0}}x\tilde{f}(x,Q^2)
\end{equation}
where 
\begin{equation}
\label{E19}
\tilde{D_0}=\sum \tilde{D}_0^i
\end{equation}
For very small $x$ and large $Q^2$, the second term of Eqn(\ref{E18}) can be neglected leading to 
\begin{equation}
\label{E20}
\tilde{q}_i (x,Q^2)= \frac{e^{\tilde{D}_0^i} Q_0^2 \left(log\frac{1}{x}\right)^{\tilde{D}_2+\tilde{D}_1 log\left(1+\frac{Q^2}{\tilde{Q}_0^2}\right)}}{M^2 \left(1+\tilde{D}_3+\tilde{D}_1 log log\left(\frac{1}{x}\right)\right)} \left(1+\frac{Q^2}{\tilde{Q}_0^2}\right)^{\tilde{D}_3 +1}
\end{equation}
which satisfies the Froissart saturation condition, if 
\begin{equation}
\label{E21}
\tilde{D}_2+ \tilde{D}_1 log\left( 1+ \frac{Q^2}{Q_0^2}\right)=2
\end{equation}
within $log log\left(\frac{1}{x}\right)$ corrections \cite{DK7}.

\section{Results}
\label{C}
\subsection{Analysis of singularity free model}
To determine the parameters of the model $\left( D_0,\ D_1,\ D_2,\ D_3,\ Q_0^2\right)$ we use recently compiled HERA data \cite{HERA} instead of earlier data \cite{H1,ZE} used in Ref\cite{Last}. Following this procedure of Ref\cite{Last} we make $\chi^2$-analysis of the data and find the following results.\\
\\
\textbf{Case 1 :} We note that $D_2$=1 is ruled out since it will make the Structure Function Eqn(\ref{E10}) $x$-independent. In Table \ref{Table1} we show the results. From the $\chi^2$-analysis, the model in case 1 is confined well with data for 0.35$\leq Q^2\leq$70GeV$^2$ and 6.62$\times$ 10$^{-6}\leq$ x $\leq$ 0.08. Note that $D_3$ and $D_1$ are taken to be zero in this limit. Here the number of $F_2$ data points is 222.\\
\begin{table}[h]
\caption{Results of the fit}
\label{Table1}
\begin{center}
\begin{tabularx}{12cm}{|X|X|X|X|X|X|X|}
\hline
$D_0$ & $D_1$ & $D_2$ & $D_3$ & $Q_0^2$ (GeV$^2$) & $\chi^2$ &$\chi^2/ndf$ \\ \hline
-4.129 \tiny$\pm0.332$  & 0 & 1.226 \tiny$\pm0.01$  & 0 & - & 180.11 & 0.81 \\ \hline
\end{tabularx}

\end{center}
\end{table}\\ \\
\\
\textbf{Case 2 :} The parameters $D_0,\ D_2,\ D_3$ and $Q_0^2$ are determined (given in Table \ref{Table2}) in the similar manner as in case 1 but the validity range is new 0.35 $\leq Q^2\leq 27$ GeV$^2$ and 6.62$\times$ 10$^{-6}\leq$ x $\leq$ 0.032. As in case 1, here too $D_2=1$ is ruled out since it will make Eqn(\ref{E12}) $x$-independent. The number of $F_2$ data points is 174.\\
\begin{table}[h]
\caption{Results of the fit}
\label{Table2}
\begin{center}
\begin{tabularx}{12cm}{|X|X|X|X|X|X|X|}
\hline
$D_0$ & $D_1$ & $D_2$ & $D_3$ & $Q_0^2$ (GeV$^2$) & $\chi^2$ &$\chi^2/ndf$ \\ \hline
-6.125 \tiny$\pm0.444$ & 0 & 1.214 \tiny$\pm0.01$ & 0.531 \tiny$\pm0.01$ & 0.053 \tiny$\pm0.001$ & 138.67 & 0.80 \\ \hline
\end{tabularx}

\end{center}
\end{table}\\ \\
\textbf{Case 3 :} Here parameters are best fitted in the range 0.35$\leq Q^2\leq$15 GeV$^2$ and 6.62$\times$ 10$^{-6}\leq$ x $\leq$ 0.02 and given in Table \ref{Table3}. The number of $F_2$ data points is 146.\\
\begin{table}[h]
\caption{Results of the fit}
\label{Table3}
\begin{center}
\begin{tabularx}{12cm}{|X|X|X|X|X|X|X|}
\hline
$D_0$ & $D_1$ & $D_2$ & $D_3$ & $Q_0^2$ (GeV$^2)$ & $\chi^2$ &$\chi^2/ndf$ \\ \hline
-3.533 \tiny$\pm0.350$ & 0.411 \tiny$\pm0.02$ & 0.582 \tiny$\pm0.003$ & 0 & 0.035 \tiny$\pm0.0005$ & 117.31 & 0.80 \\ \hline
\end{tabularx}

\end{center}
\end{table}\\ \\
\textbf{Case 4 :} Parameters $D_0,\ D_1,\ D_2,\ D_3$ and $Q_0^2$ are determined and given in Table \ref{Table4} and obtained in a more restrictive range 0.85$\leq Q^2\leq$10 GeV$^2$ and 2$\times$ 10$^{-5}\leq$ x $\leq$0.02. The number of $F_2$ data points is 95. \\
\begin{table}[h]
\caption{Results of the fit}
\label{Table4}
\begin{center}
\begin{tabularx}{12cm}{|X|X|X|X|X|X|X|}
\hline
$D_0$ & $D_1$ & $D_2$ & $D_3$ & $Q_0^2$ (GeV$^2)$ & $\chi^2$ &$\chi^2/ndf$ \\ \hline
-2.971 \tiny$\pm0.409$ & 0.065 \tiny$\pm0.0003$ & 1.021 \tiny$\pm0.004$ & 0.0003 & 0.20 \tiny$\pm0.0008$ & 18.829 & 0.20 \\ \hline
\end{tabularx}

\end{center}
\end{table}\\

\begin{figure*}
\centering
\includegraphics[width=0.79\textwidth]{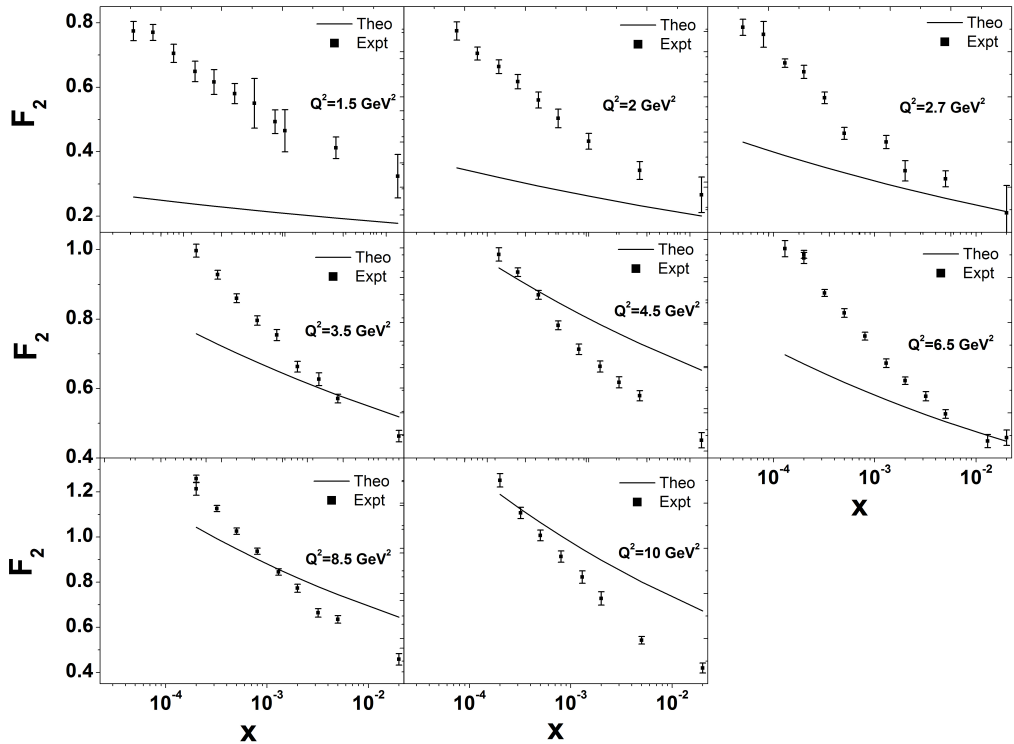}
\caption{Measurement of the structure function $F_2(x,Q^2)$ as a function of $x$ in bins of $Q^2$ by recently compiled HERA data.}
\label{Fig1}
\end{figure*}

In Figure \ref{Fig1},  we plot $F_2$ as a function of $x$ for eight representative values of $Q^2$ ($Q^2$= 1.5, 2, 2.7, 3.5, 4.5, 6.5, 8.5, 10 GeV$^2$) in the phenomenological allowed range 0.85$\leq Q^2\leq$10 GeV$^2$. We also show the corresponding available data from Ref \cite{HERA}.

Our analysis indicates that the phenomenological range of validity of the present version of the model is more restrictive $0.85\leq Q^2 \leq 10$ GeV$^2$ and $2\times 10^{-5} \leq x \leq 0.02$ to be compared with Eqn(\ref{E8}) of the previous version of Ref\cite{Last}. 

We also observe the following features of the model compared to data: at $Q^2=1.5$GeV$^2$ data overshoots the theory. But as $Q^2$ increases, the theoretical curve comes closer to data. At $Q^2$=10 GeV$^2$, on the other hand, the theory exceeds data. Main reason of this feature is that the $x$-slope of the model is less than that of the data. This limitation can in principle be overcome by modification of the magnification factor $\frac{1}{x}$ to $\left(\frac{1}{x}-1\right)$ in Eqn(\ref{E1}) to accommodate proper large $x$ behavior of structure function as noted in Ref\cite{DK6}.

\begin{figure}[h]
\centering
\includegraphics*[width=62mm]{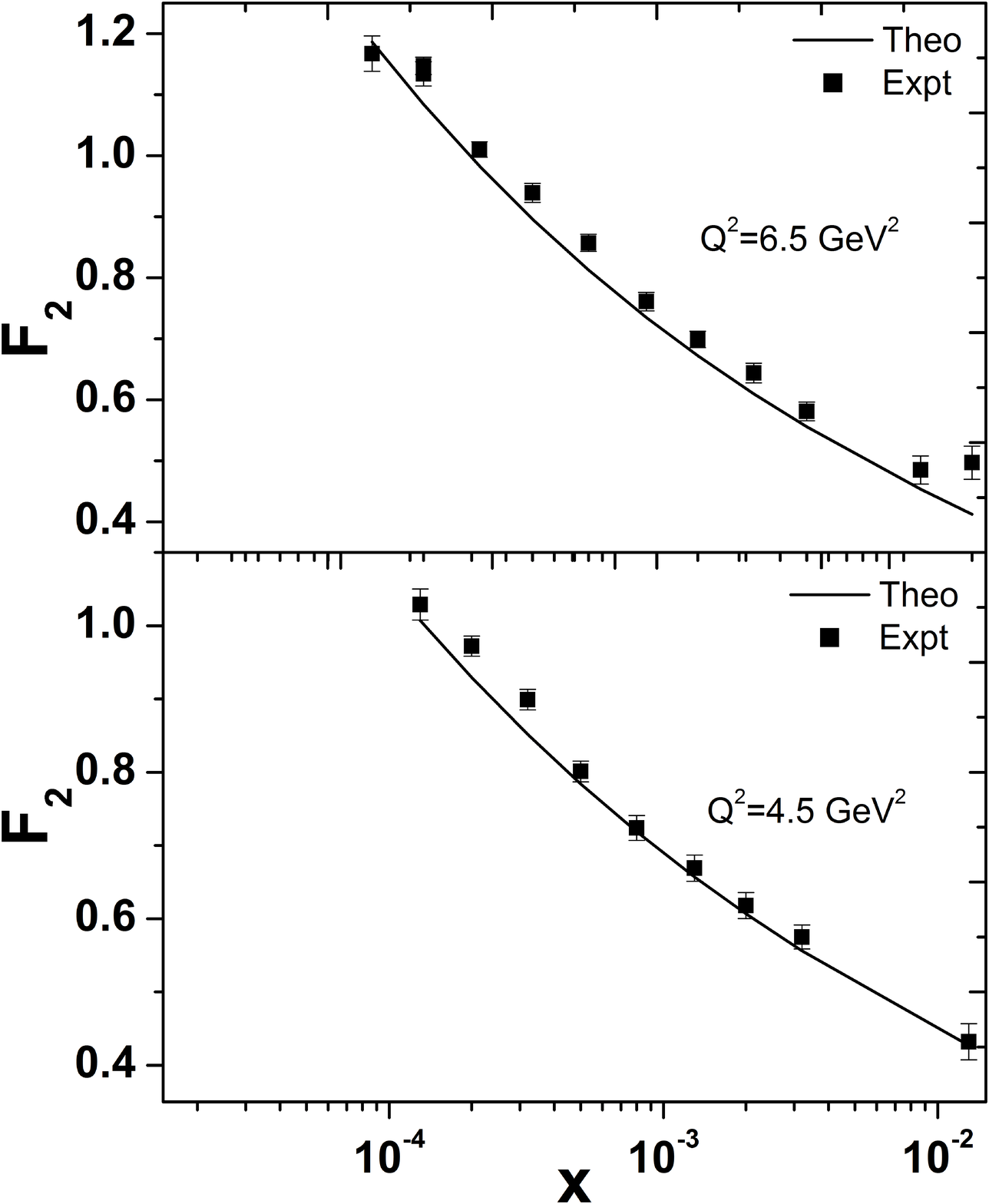}
\caption{Measurement of the structure function $F_2(x,Q^2)$ as a function of $x$ in bins of $Q^2$ by recently compiled HERA data.}
\label{Fig2}
\end{figure}

As an illustration we show results of two representative values of $Q^2$= 4.5 and 6.5 GeV$^2$ in Figure~\ref{Fig2} with such magnification using the parameters in Table~\ref{Table4}.

\subsection{Graphical representation of TMD}
It is interesting to predict the $k_t^2$-dependance of unintegrated parton distribution (TMD) from the $x$ and $Q^2$ dependence of the integrated parton distribution function (PDF). Clearly this can be done within a model framework, as has been noted in Ref\cite{ZA} as well as in Ref\cite{DK9}. Though it should be of interest to explore this approach to study $k_t^2$-dependence of $f_i(x,k_t^2)$, such a study makes sense only in the $x$-$Q^2$ range where the approach works and where the parameters had been fitted namely $2\times 10^{-5} \leq x \leq 0.02$ in the present model. 

\begin{figure*}
\captionsetup[subfigure]{labelformat=empty}
\centering
\subfloat[]{\includegraphics[width=.4\textwidth]{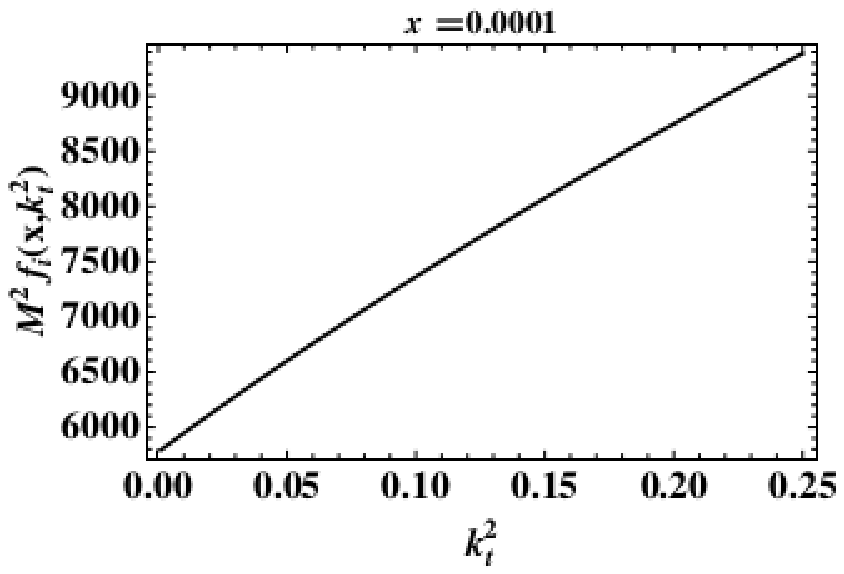}}\quad
\subfloat[]{\includegraphics[width=.4\textwidth]{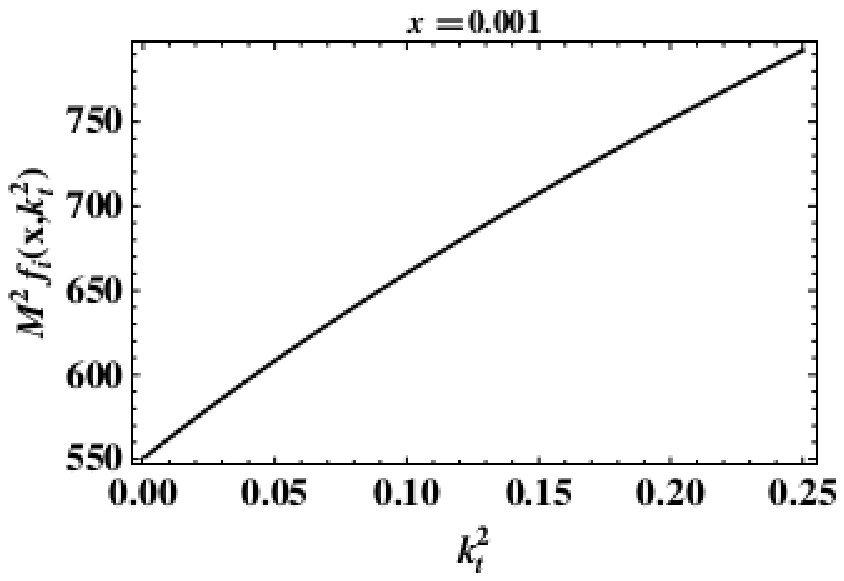}}\quad
\subfloat[]{\includegraphics[width=.4\textwidth]{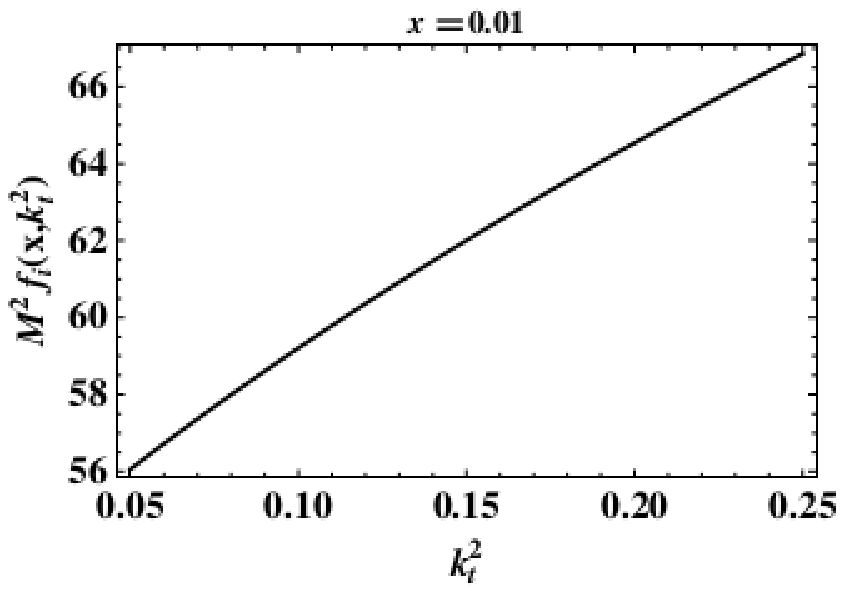}}\quad
\subfloat[]{\includegraphics[width=.4\textwidth]{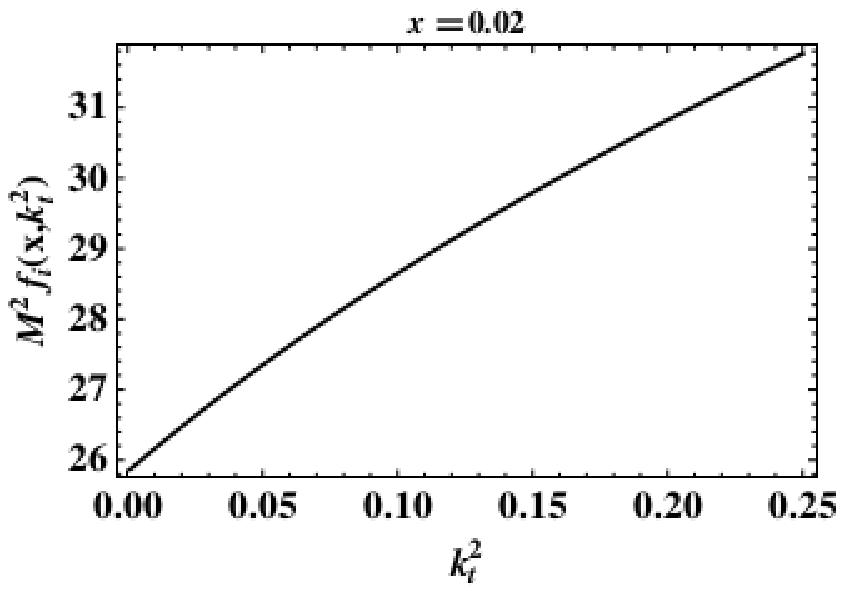}}\quad
\caption{$M^2 f_i(x,k_t^2)$ vs $k_t^2$}
\label{Fig3}
\end{figure*}

Using Eqn(\ref{E1}), $f_i(x,k_t^2)$ has the form
\begin{equation}
\label{E22}
f_i(x,k_t^2)=\frac{e^{D_0^i}}{M^2} \left(\frac{1}{x}\right)^{D_2+D_1 log\left(1+\frac{k_t^2}{Q_0^2}\right)} \left(1+\frac{k_t^2}{Q_0^2}\right)^{D_3}
\end{equation}\\
which has dimension of mass$^{-2}$ consistence with Eqn(\ref{E2}) where the PDF $q_i(x,Q^2)$ is dimensionless. To evaluate Eqn(\ref{E22}) we take the mean value $D_0$, $D_1$, $D_2$, $D_3$ and $Q_0^2$ from the Table\ref{Table4}. Assuming $D_0=n_f D_0^i$ and setting $n_f=4$, we obtain $D_0^i= -0.742$. In Figure~\ref{Fig3}, $M^2 f_i(x,k_t^2)$ vs $k_t^2$ is shown for representative values of \\
(i) $x$ = 10$^{-4}$ (ii) $x$ = 10$^{-3}$ (iii) $x$ = 10$^{-2}$ (iv) $x$=0.02 setting $M^2=1$ GeV$^2$.

The allowed limit of $k_t^2$ is considered to be less than the average value $k_t^2$=0.25 GeV$^2$ \cite{h2} as determined from data.

\begin{figure*}
\captionsetup[subfigure]{labelformat=empty}
\centering
\subfloat[]{\includegraphics[width=.4\textwidth]{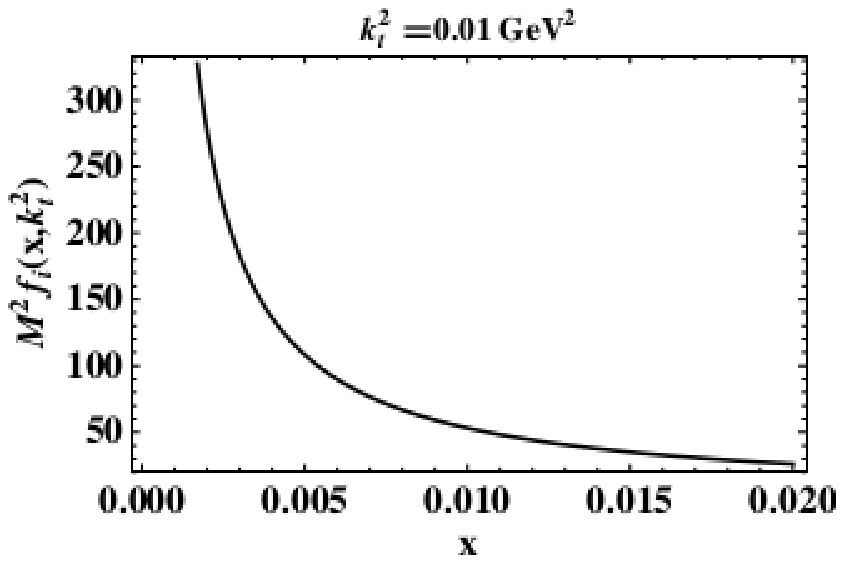}}\quad
\subfloat[]{\includegraphics[width=.4\textwidth]{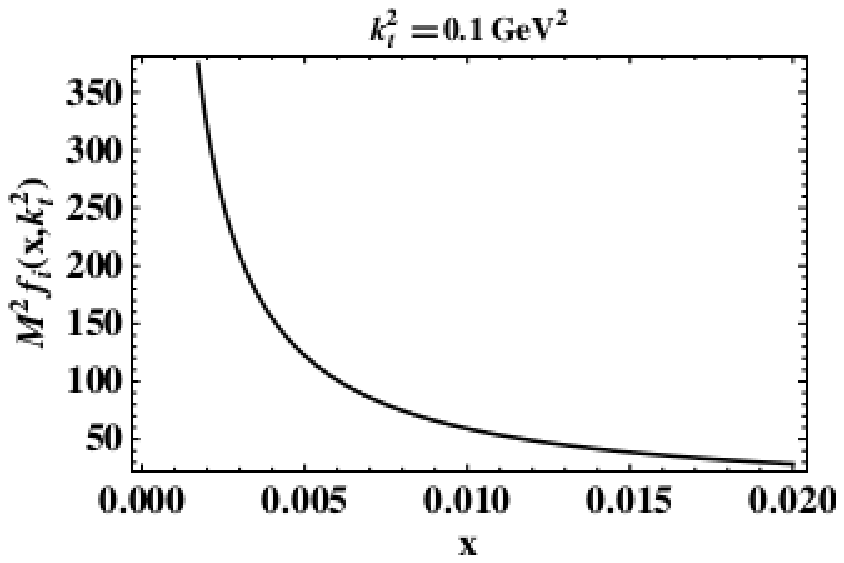}}\quad
\subfloat[]{\includegraphics[width=.4\textwidth]{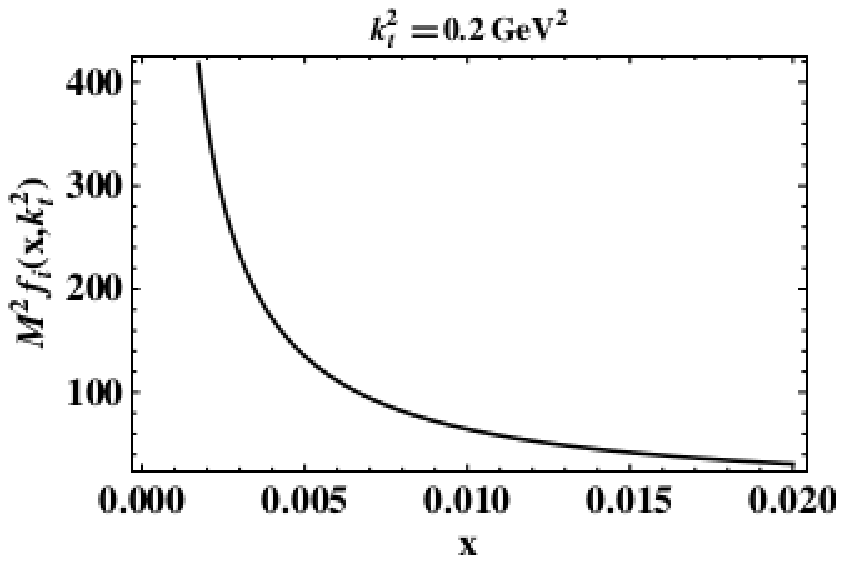}}\quad
\subfloat[]{\includegraphics[width=.4\textwidth]{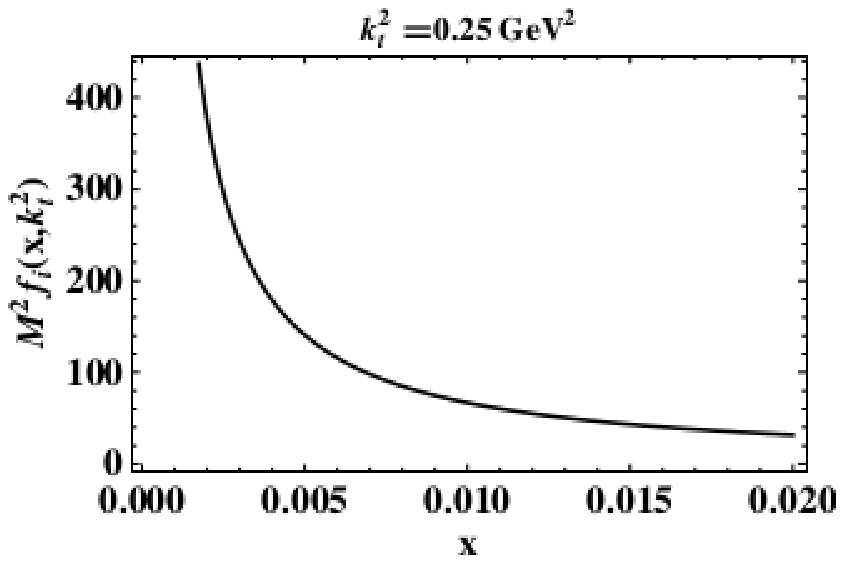}}\quad
\caption{$M^2 f_i(x,k_t^2)$ vs $x$}
\label{Fig4}
\end{figure*}

Similarly $M^2 f_i(x,k_t^2)$ vs $x$ as shown in Figure~\ref{Fig4} for representative values of \\ 
(i) $k_t^2$ = 0.01 GeV$^2$ (ii) $k_t^2$ = 0.1 GeV$^2$ (iii) $k_t^2$= $Q_0^2$ = 0.20 GeV$^2$
(iv) $k_t^2$=0.25 GeV$^2$.

The present graphical analysis of TMD [Fig~\ref{Fig3},\ref{Fig4}] is an improvement over the earlier one \cite{DK9}, in the sense that the experimental limit on $k_t^2$ as determined from data \cite{h2} was not taken into consideration there. Further it was also beyond the range of validity of the original model. Proper dimension of the TMD has also been taken into account in the present case.\\

Let us now compare the structure of the model TMD (Eqn\ref{E22}) with the suggested forms available in current literature.

\begin{figure}
\centering
\includegraphics*[width=70mm]{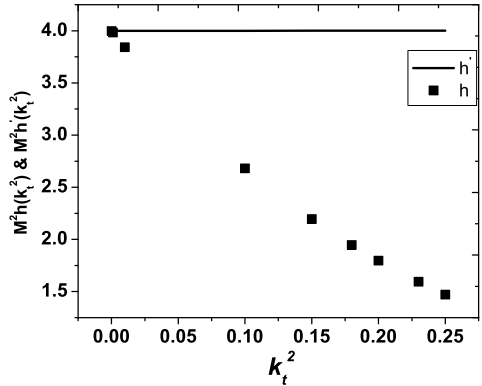}
\caption{$M^2 h(k_t^2)$  $M^2 h'(k_t^2)$ vs $k_t^2$}
\label{Fig5}
\end{figure}

The standard way to study TMDs is through the factorization approach \cite{h1,h2,holo} where $x$ and $Q^2$-dependence are factorized into a PDF $q_i(x,Q^2)$ and a Gaussian transverse momentum dependent function $h(k_t^2)$ 
\begin{equation}
\label{E23}
f_i(x,k_t^2,Q^2)= q_i(x,Q^2) h(k_t^2)
\end{equation}
where, 
\begin{equation}
\label{E24}
h(k_t^2)= \frac{1}{\langle k_t^2\rangle} e^{-\frac{k_t^2}{\langle k_t^2\rangle}}
\end{equation}
with normalization condition, 
\begin{equation}
\label{E25}
\int h(k_t^2)dk_t^2= 1
\end{equation}\\
Such factorization property of TMD is not present in the model, (Eqn\ref{E22}) nor the Gaussian form Eqn\ref{E24}. Only in the absence of correlation term $D_1$ (Eqn\ref{E1}) such factorization property emerges. In this limit the $k_t^2$ dependent functional form of the TMD is given by 
\begin{equation}
\label{E26}
h'(k_t^2)= \frac{1}{M^2}\left(1+\frac{k_t^2}{Q_0^2}\right)^{D_3}
\end{equation}
in contrast to a Gaussian function (Eqn\ref{E24}). Introducing a $k_t^2$ cut off \ $0<k_t^2<\langle k_t^2\rangle$ \ Eqn(\ref{E26}) will satisfy the normalization condition (Eqn\ref{E25}) with a normalization constant 
\begin{equation}
\label{E27}
N= \frac{M^2 (D_3+1)}{Q_0^2\left[\left(1+\frac{\langle k_t^2\rangle}{Q_0^2}\right)^{D_3+1}-1\right]} 
\end{equation}\\
In figure~\ref{Fig5} we compare Eqn(\ref{E24}) and Eqn(\ref{E26}) with $\langle k_t^2\rangle$= 0.25 GeV$^2$ in $M^2 h(k_t^2)$ $M^2 h'(k_t^2)$ vs $k_t^2$ plot.

\subsection{Possibility of Froissart saturation in the model}
The condition of Froissart saturation in the parton distribution function in the model \cite{DK7} leads to modification of the magnification factor $\frac{1}{x}$ to $log\frac{1}{x}$ with the additional condition Eqn(\ref{E21}).

For fixed $\tilde{D_1}$ and $\tilde{D_2}$ it can be satisfied only for a specific value of $Q^2$. As an illustration if one takes the mean values of the set of Table \ref{Table4}, i.e. $\tilde{D_0}\approx$ -2.971 GeV$^2$, $\tilde{D_1}\approx 0.065$GeV$^2$, $\tilde{D_2}\approx$ 1.021 GeV$^2$, $\tilde{D_3}\approx$ 0.0003 GeV$^2$, $Q_0^2\approx$ 0.20 GeV$^2$ it is obtained at $Q_s^2$ =6.95$\times 10^{5}$ GeV$^2$, far beyond the phenomenological range of validity 0.85$\leq Q^2 \leq$ 10 GeV$^2$. However its leading $log^2\frac{1}{x}$ behavior is to be manifested to any $Q^2$, $Q^2 \geq Q_s^2$ or in a specific regime of $Q^2$, 0.11$< Q^2 < $1200 GeV$^2$ as in Ref\cite{BL} $D_1$ and $D_2$ should have proper $Q^2$-dependence, a feature beyond the scope of the present method of parametrization.

A leading $log^2\frac{1}{x}$ behavior of the structure function $F_2(x,Q^2)$ with $Q^2$-independent exponent is possible in the present model only if the correlation term proportional to $D_1$ in the definition of TMD (Eqn\ref{E1}) is negligible compared to $D_2$ and $D_3$ and it is redefined as 
\begin{equation}
\label{E28}
x f_i(x, k_t^2)= \frac{e^{D_0^i}}{M^2} \left(1+\frac{k_t^2}{Q_0^2}\right)^{D_3} \left(log\frac{1}{x}\right)^{D_2}
\end{equation}

Instead of Eqn(\ref{E22}), in such a case the corresponding PDF and structure function have form as
\begin{equation}
\label{E29}
x q_i(x,Q^2)= \frac{e^{D_0^i}}{M^2} \frac{\left(log\frac{1}{x}\right)^{D_2}}{1+D_3} \left( \left(1+\frac{Q^2}{Q_0^2}\right)^{D_3+1}-1\right)
\end{equation}

and
\begin{equation}
\label{E30}
F_2(x,Q^2)= \frac{e^{D_0}}{M^2} \frac{\left(log\frac{1}{x}\right)^{D_2}}{1+D_3} \left( \left(1+\frac{Q^2}{Q_0^2}\right)^{D_3+1}-1\right)
\end{equation}

Note that Eqn(\ref{E29}) and Eqn(\ref{E30}) corresponds to case 2 (Eqn\ref{E11} and Eqn\ref{E12}) of the original model.

Setting $D_2$= 2 Eqn(\ref{E28}-\ref{E30}) will then yield the desired Froissart saturation saturation in the model. It is to be noted that condition (\ref{E21}) [Eqn(28) of Ref\cite{DK7}] corresponds to the Froissart compatibility of the PDF and not the structure finction $F_2(x,Q^2)$ as defined in Eqn(\ref{E5}) due to the multiplicative $x$-factor relative to the PDF.

Redefining of TMD in Eqn(\ref{E28}) resolves this anomaly.

\section{Summary}
\label{D}
In the present paper, we have made a reanalysis of a structure function $F_2^p(x,Q^2)$ based on self-similarity \cite{Last}. The present study is based on the notion that a physically viable model proton should be finite in the $x$-range $0< x <1$ hence singularity free. It also conforms to the expectation that $``$fractal dimension" associated with self-similarity are invariably positive definite. We have then studied the possibility of incorporating Froissart saturation bound.

Our analysis indicates that the range of validity of the present version of the model approaches non-perturbative regime of lower $Q^2$, \ $0.85\leq Q^2\leq 10$GeV$^2$ instead of $0.045\leq Q^2\leq 120$GeV$^2$ of the Ref\cite{Last}. It is also suggested that the $x$-slope of the predicted structure function can be increased by proper redefinition of the defining magnification factors $\frac{1}{x}$ as noted in Ref\cite{DK6}. Interestingly, the range of validity of the present version of the model is close to that of Ref\cite{holo} based on holographic QCD.

Pattern of momentum fractions carried by the quarks and gluons in the present model [case 1-4] is currently under study. 

\section*{Acknowledgment}
We thank Dr. Kushal Kalita for helpful discussions, Dr. Rupjyoti Gogoi of Tezpur University for collaboration at the initial stage of the work and Dr. Akbari Jahan for useful comments on self-similarity. One of the authors (B.S.) acknowledges the UGC-RFSMS for financial support.


\begin{thebibliography}{10}

%\section*{Reference}

\bibitem{Last} T. Lastovicka, Euro. Phys. J. C \textbf{24}, 529 (2002), hep-ph/0203260
\bibitem{DK1} D.K. Choudhury and Rupjyoti Gogoi, hep-ph/0310260; hep-ph/0503047
\bibitem{DK2} D.K. Choudhury and Rupjyoti Gogoi, Indian. J. Phys. \textbf{80}, 823 (2006)
\bibitem{DK3} D.K. Choudhury and Rupjyoti Gogoi, Indian. J. Phys.\textbf{81}, 607 (2007)
\bibitem{DK4} A. Jahan and D.K. Choudhury, Mod. Phys. Lett. A \textbf{27}, 1250193 (2012), hep-ph/1304.6882
\bibitem{DK5} A. Jahan and D.K. Choudhury, Mod. Phys. Lett. A \textbf{28}, 1350056 (2013), hep-ph/1306.1891
\bibitem{DK6} D.K. Choudhury and A. Jahan, Int. J. Mod.Phys. A \textbf{28}, 1350079 (2013), hep-ph/1305.6180
\bibitem{DK7} A. Jahan and D.K. Choudhury, Phys. Rev. D \textbf{89}, 014014 (2014), hep-ph/1401.4327
\bibitem{DK8} A. Jahan and D.K. Choudhury, hep-ph/1404.0808
\bibitem{H1} H1:C. Adloff \textit{et al.}, Euro. Phys. J. C \textbf{21}, 33-61 (2001), hep-ex/0012053
\bibitem{ZE} ZEUS: J. Breitweg \textit{et al.}, Phys. Lett. B \textbf{487}, 53 (2000), hep-ex/0005018
\bibitem{GRV} M. Gl{\"u}ck, E Reya and A. Vogt, Euro. Phys. J. C \textbf{5}, 461 (1998)
\bibitem{BL} M.M. Block, L. Durand, P. Ha and D.W. McKay, Phys. Rev. D \textbf{84}, 094010 (2011) 
\bibitem{FR} F. Froissart Phys. Rev. \textbf{123}, 1053 (1961) 
\bibitem{ROY} A. Martin and  S.M. Roy, hep-ph/1306.5210
\bibitem{HERA} H1 and ZEUS Collaborations, JHEP \textbf{01}, 109 (2010)
\bibitem{ZA} P. Zavada, Phys. Rev. D \textbf{83}, 014022 (2011), hep-ph/0908.2316
\bibitem{DK9} A. Jahan and D.K. Choudhury, hep-ph/1106.1145 
\bibitem{h} M. Anselmino \textit{et al.}, Phys. Rev. D \textbf{71}, 074006 (2005), hep-ph/0501196
\bibitem{h1} M. Anselmino \textit{et al.}, Phys. Rev. D \textbf{75}, 054032 (2007), hep-ph/0707.1197
\bibitem{h2} M. Anselmino \textit{et al.}, Nucl. Phys. Proc. Suppl. \textbf{191}, 98-107 (2009), hep-ph/0812.4366
\bibitem{holo} Akira Watanabe and Katsuhiko Suzuki, hep-ph/1312.7114
\end{thebibliography}
\end{document}